\documentstyle [12pt] {article}

\catcode`@=12
\begin{document}
\baselineskip 24pt
\thispagestyle{empty}
\begin{flushright} UCRHEP-T102\\November 1992\
\end{flushright}
\vspace {0.5in}
\begin{center}

{\Large \bf Hints within the Standard Model on $m_t$ and $m_H$\\}
\vspace{1.8in}
{\bf Ernest Ma\\}
\vspace{0.3in}
{\sl Department of Physics\\}
{\sl University of California\\}
{\sl Riverside, California 92521\\}
\vspace{1.2in}
\end{center}
\begin{abstract}\
If one or more otherwise divergent quantities in the
standard model are actually finite, they may be indications of underlying
dynamics.  In particular, one-loop finiteness of the $m_H$ renormalization
is achieved if $m_t^2 \simeq m_H^2 = (2M_W^2 + M_Z^2) / 3$.
\end{abstract}
\vspace{0.6in}
*To appear in Proc. of DPF92 (Fermilab, Nov 1992).

\newpage
\section{Introduction}
	The standard model has a quadratic divergence proportional to
\begin{equation}
2\lambda + {1 \over 2} g_1^2 + {3 \over 2} g_2^2 - 4 \sum_f \left( {n_f \over
3} \right) g_f^2,
\end{equation}
where $\lambda$ is the quartic scalar self-coupling, $g_1$ the U(1) gauge
coupling, $g_2$ the SU(2) gauge coupling, and $g_f$ the Yukawa couplings of
the fermions $f$ to the Higgs boson, with $n_f$ the number of colors,
{\it i.e.} 3 for quarks and 1 for leptons.  This may very well be just an
artifact of the regularization procedure and we can forget all about it after
proper renormalization of all the physical quantities.  Alternatively, we
may take it seriously as a hint that new physics will come in at some energy
scale higher than the electroweak scale of $10^2$ GeV and make it finite.
In a scenario involving supersymmetry, new particles will appear below 1 TeV
or so and cancel the divergence associated with each term of the above
expression.  In a scenario without new particles up to an energy scale
$\Lambda >>$ 1 TeV, it may be conjectured that whatever the underlying
dynamics, it should be such that the above expression is suppressed,
say of order ($10^2$GeV/$\Lambda)^2$, which must then come about
from the cancellation among the various couplings.

	Attaching $v^2$ (square of the Higgs-boson vacuum expectation value)
to (1) and setting it equal to zero, we obtain the well-known Veltman
condition\cite{veltman}
\begin{equation}
4m_t^2 \simeq 2M_W^2 + M_Z^2 + m_H^2,
\end{equation}
where all other fermion masses have been dropped because their contributions
are negligible.  This condition is consistent with the present experimental
data $M_Z = 91.175 \pm 0.021$ GeV, $M_W = 80.14 \pm 0.27$ GeV, $m_t > 91$ GeV,
and $m_H > 60$ GeV.

\section{A Closer Look}
	Since $\lambda$, $g_1^2$, $g_2^2$, and $g_f^2$ change as functions of
$q^2$, and the expression (1) is not invariant under this change, the Veltman
condition (2) should apply only at one unique mass scale.  In the standard
model, the presence of spontaneous symmetry breaking implies the existence
of a tadpole diagram in which a physical Higgs boson ends up in a loop
involving all the massive particles.  This diagram is quadratically
divergent and contributes to all self-masses.  Hence the natural choice is
$q^2 = m_H^2$.

\section{Recent Conjectures}
	In addition to the condition (2), is there another hint within the
standard model of a possible relationship among couplings?  Perhaps a
particular logarithmic divergence should be suppressed as well.  This is
the essence of 3 recent conjectures.

	Osland and Wu\cite{oslandwu} singled out the $He^+e^-$ coupling and
required its logarithmic divergence to be zero.  This results in the condition
\begin{equation}
m_t^2 \simeq {5 \over 2} M_Z^2 - M_W^2.
\end{equation}
However it is not clear why this particular coupling should be chosen instead
of some other, and once it is chosen, we must still define it at some $q^2$
because this logarithmic divergence cannot be zero at all mass scales.

	Blumhofer and Stech\cite{blumstech} proposed to set the logarithmic
divergence of the Higgs tadpole also to zero.  This has the advantage of
a well-defined $q^2$, {\it i.e.} $m_H^2$, but the procedure is
gauge-dependent and therefore suspect.  However, they argued that the
choice $\xi=0$ in the $R_\xi$ gauge would correspond to a gauge-invariant
physical quantity having to do with vacuum condensates.  This then implies
\begin{equation}
4m_t^4 \simeq 2M_W^4 + M_Z^4 + {1 \over 2} m_H^4.
\end{equation}

	Decker and Pestieau\cite{deckpest} chose the mass of the electron
neutrino and required its logarithmic divergence to be zero, assuming of
course that there is a right-handed singlet partner to the observed left-handed
neutrino and they combine to allow a Dirac mass.  This results in the
following condition
\begin{equation}
4m_t^4 \simeq 2M_W^4 + M_Z^4 + {1 \over 2} m_H^4 + {1 \over 2} (m_e^2 -
m_{\nu_e}^2) m_H^2,
\end{equation}
which is almost identical to (4).  Again it is not clear why this particular
mass (which may not even exist) should be chosen instead of some other.

\section{The Most Natural Choice}
	If a particular logarithmic divergence is to be chosen zero in
addition to the Veltman condition, the most natural choice is clearly that of
the Higgs-boson mass itself.\cite{ma}  After all, it is uniquely defined
at $q^2=m_H^2$ as already assumed in (2).  It is also gauge-independent.
The resulting condition is
\begin{equation}
2m_t^2 \simeq 2M_W^2 + M_Z^2 - m_H^2,
\end{equation}
which, when combined with (2), implies
\begin{equation}
m_t^2 \simeq m_H^2 = {2 \over 3} M_W^2 + {1 \over 3} M_Z^2.
\end{equation}
If dimensional regularization is used to extract the quadratic divergence of
the standard model, the residue of the pole at $d=2$ depends also on $d$ and
the Dirac trace.  To get the Veltman condition, we have to set both equal to
4.  Perhaps we should\cite{oslandwu} really use the value 2, then instead of
(2), we find
\begin{equation}
6m_t^2 \simeq 2M_W^2 + M_Z^2 + 3m_H^2.
\end{equation}
Remarkably, when combined with (6), the condition (7) is again obtained.
Hence the proposed conjecture of one-loop finite $m_H$ renormalization is
independent of the regularization procedure for the quadratic divergence.

\section{Conclusion}
	Numerically, the condition (7) implies
\begin{eqnarray}
m_t & \simeq & 84~{\rm GeV~+~higher-order~corrections}, \\
m_H & \simeq & 84~{\rm GeV~+~higher-order~corrections},
\end{eqnarray}
whereas present data require $m_t > 91$ GeV, and $m_H > 60$ GeV.  Hence the
above conjecture is on the verge of being ruled out.  On the other hand,
if either (2) or (6) turns out to be approximately satisfied, it may still
be an indication of underlying dynamics.

	It should be noted that the above conditions are all based on
only one-loop contributions and there is no explicit reference to the
mass scale $\Lambda$ of new physics.  The higher-order contributions, all
defined at $q^2=m_H^2$, are considered as small corrections, but they will
depend on $\Lambda$ logarithmically.

\begin{center} {ACKNOWLEDGEMENT}
\end{center}
	This work was supported in part by the U. S. Department of Energy
under Contract No. DE-AT03-87ER40327.

\newpage
\bibliographystyle{unsrt}

\end{document}